\documentclass[aps,prd,nofootinbib,nobibnotes,notitlepage,twocolumn]{revtex4-2}
\usepackage{graphicx}
\usepackage{amssymb}
\usepackage{amsmath} 
\usepackage{color}
\usepackage{float}
\usepackage{ulem}
\usepackage{accents}

\usepackage{amsfonts}
\usepackage[colorlinks=true,
pdfstartview=FitV,linkcolor=blue,
citecolor=blue,urlcolor=blue,breaklinks=true]{hyperref}
\usepackage{array}
\usepackage{float}
\usepackage{placeins}
\usepackage[dvipsnames]{xcolor}
\usepackage{csquotes}
\usepackage[makeroom]{cancel}
\usepackage{units}
\usepackage{fixmath}
\usepackage{bigints}
\usepackage{amssymb}
\usepackage{systeme}
\newcolumntype{C}[1]{>{\centering\arraybackslash}m{#1}}
\renewcommand{\eqref}[1]{\mbox{Eq.~(\ref{#1})}}

\definecolor{ForestGreen}{rgb}{0.13,0.55,0.13}

\makeatletter
\renewcommand*\l@section{\@dottedtocline{1}{0em}{1.5em}}
\renewcommand*\l@subsection{\@dottedtocline{1}{1.5em}{1.5em}}
\renewcommand*\l@subsubsection{\@dottedtocline{1}{3em}{1.5em}}
\makeatother

\begin{document}

\title{Anomalous increasing super reflectance in chiral matter}

\author{Pedro D. S. Silva$^{a}$}
\email{pedro.dss@ufma.br} \email{pdiegoss.10@gmail.com}
\author{Alex Q. Costa$^b$}
\email{costa.alex@discente.ufma.br, prof.costalex@gmail.com}
\author{Ronald A. Pereira$^{b}$}\email{ronald.ap@discente.ufma.br}\email{ronald123.araujo@gmail.com}
\author{Manoel M. Ferreira Jr.$^{b,c}$}
\email{manojr.ufma@gmail.com; manoel.messias@ufma.br}
	\affiliation{$^a$Coordena\c{c}\~ao do Curso de Ci\^encias Naturais - F\'isica, Universidade Federal do Maranh\~ao, Campus de Bacabal, Bacabal, Maranh\~ao, 65700-000, Brazil}
\affiliation{$^b$Programa de P\'{o}s-gradua\c{c}\~{a}o em F\'{i}sica, Universidade Federal do Maranh\~{a}o, Campus Universit\'{a}rio do Bacanga, S\~{a}o Lu\'is (MA), 65080-805, Brazil}
\affiliation{$^c$Departamento de F\'{\i}sica, Universidade Federal do Maranh\~{a}o,
Campus Universit\'{a}rio do Bacanga, S\~{a}o Lu\'is (MA), 65080-805, Brazil}

\begin{abstract}

Magnetic and anomalous Hall conductivities induce anomalous transport features and novel optical phenomena in chiral systems. Here, we investigate reflection properties on the surface of a medium ruled by axion electrodynamics, which effectively describes optical aspects of Weyl semimetals. We show that these chiral media can manifest anomalous reflectance (R greater than unity) for some frequency windows, depending on the signs of the two involved conductivities. Such a reflectance can increase with the frequency, being always greater than 1 in certain frequency bands. We also examine the complex Kerr angles at normal incidence on the chiral medium. Giant Kerr angle is observed within the absorption window, while the Kerr ellipticity may be used to determine the relative sign of the magnetic conductivity.

\end{abstract}


\maketitle


\textbf{\textit{Introduction}.} Anomalous transport phenomena have been extensively investigated in a class of topological physical systems, such as the Weyl semimetals \cite{Armitage}, where an electric current,
\begin{equation}
\mathbf{J}_{CME}=\sigma^B {\bf{B}}, \label{CMEcurrent1}
\end{equation}
with $\sigma^B =e^2 \mu_{5}/(2\pi^2\hbar^2)$, is generated along an applied magnetic field due to the axial chemical potential, $\mu_{5}=\mu_{R}-\mu_{L}$, that reflects an imbalance in the number density of chiral fermions with opposite handedness. This current is a signature of the chiral magnetic effect (CME)\cite{Kharzeev1, Fukushima}, which has been widely addressed in the study of condensed matter physics \cite{Burkov,MChang}, relativistic plasmas \cite{Dvornikov,Schober,Wang1}, chiral magnetic instability in neutron stars \cite{Dvornikov1,Sigl}, cosmology and astrophysics \cite{Kamada}, among others. Possible signals of this effect have been detected in chiral quark-gluon plasma \cite{Belmoont} and in the Weyl semimetal samples of $ZrTe_{5}$ \cite{Li} and TaAs \cite{Xiaochun-Huang}, where it was observed a chiral-anomaly conductivity depending on $\mathbf{B}^2$.

When the time reversal symmetry is violated, the anomalous Hall effect (AHE) can occur in Weyl semimetals \cite{Nagaosa}, consisting of an electric current orthogonal to an applied electric field (in the absence of a magnetic field), 
\begin{equation}
\mathbf{J}_{AHE}=\mathbf{b} \times {\bf{E}}, \label{AHEcurrent1}
\end{equation}
where $\mathbf{b}= e^3 \mathbf{A}_{5}/(2\pi^2\hbar^2)$ and $2e\mathbf{A}_{5}$ is the momentum separation between two Weyl nodes for right-handed and left-handed fermions \cite{Zyuzin,Yang,BurkovB,Haldane,Xiao}. For a dynamical magnetic field, electromagnetic linear responses in Weyl semimetals, analyzed by the chiral kinetic theory, provided the chiral electric current (\ref{CMEcurrent1}), and, for non equilibrium configurations, unstable electromagnetic modes were also reported \cite{Nishida1}.

The AHE and the CME can be effectively described by including the axion term \cite{Qiu,Wilczek,Sekine, KDeng, Barnes}, $\mathcal{L}=\theta(\mathbf{E}\cdot\mathbf{B})$, in the Maxwell Lagrangian in continuous matter. For the special case of a nondynamical axion field, $\partial_{t}\theta=cte$, $\nabla\theta=cte$, one recovers the  Maxwell-Carroll-Field-Jackiw (MCFJ) electrodynamics \cite{CFJ,Colladay,CFJG3,Yuri,Caloni} in continuous matter \cite{Pedro2}. In this case, the Ampere's law reads $\nabla\times\mathbf{H}-\partial_{t}\mathbf{D}=\mathbf{J}+ b_{0} {\bf{B}} - \mathbf{b}\times\mathbf{E}$,
with $\partial_{t}\theta = b_{0}$ and $\mathbf{b}=\nabla\theta$ standing for the axion field time-derivative and gradient, respectively. The axion electrodynamics effectively accounts for relevant aspects of Weyl semimetals \cite{Marco}, optical properties of exotic metamaterials \cite{Barredo}, axion dielectrics \cite{Chang1,PedroPRB2024A}, Cherenkov radiation \cite{Cherenkov}, connections with the London equation and Weyl semimetals \cite{Stalhammer,Shyta}, 
optical reflection properties at the surface of an axion dielectric \cite{Pedro-Ronald-PRB2024},  applications in
the area of ultrafast magnetism \cite{Barredo2} and photonis of new chiral materials \cite{Barredo3}.

The simultaneous occurrence of AHE and CME in antiferromagnetic insulators with spin-orbit coupling was predicted as the detection of the effective axion electrodynamics in condensed matter \cite{SekineNomura,Sekine}, which has been recently observed in the context of the material $MnBi_2Te_4$ \cite{QiuGhosh}. In these new compounds, the electromagnetic propagation engenders unusual optical phenomena, that work as peculiar signatures. A relevant phenomenon is the reflected wave polarization rotation, arising when incident light is reflected on the surface of a gyrotropic medium and becomes elliptically polarized. The corresponding polarization state is given in terms of the Kerr rotation angle and Kerr ellipticity angle \cite{Sato, Argyres, Shinagawa}, used to characterize several distinct material systems, involving Weyl semimetals \cite{Kargarian, Ghosh}, topological insulators \cite{Schlenker-Souche, Sonowal}, media ruled by $CPT$-even and Lorentz symmetry violating electrodynamics \cite{Ruiz-Escobar}. Furthermore, chiral non equilibrium media may provide another non conventional feature, the anomalous reflectance, as reported below.

Parallel electric and magnetic fields applied to Weyl semimetals pump an axial charge that renders nonequilibrium steady states in the presence of both CME and AHE, whose electromagnetic behavior is marked by unstable waves at low frequencies. This instability yields anomalous reflectance ($R>1$) on the surface of the chiral medium, sustained by suplementary energy stemming from the relaxation towards the equilibrium \cite{Nishida2}. In this letter, we investigate new aspects of optical anomalous reflection at an interface between a usual dielectric and a dielectric substrate exhibiting magnetic conductivity and an anomalous Hall term, examining both the Kerr angles and the enhanced reflectance.  Throughout this work, we consider natural units.

\begin{figure}[h]
	\centering\includegraphics[scale=.27]{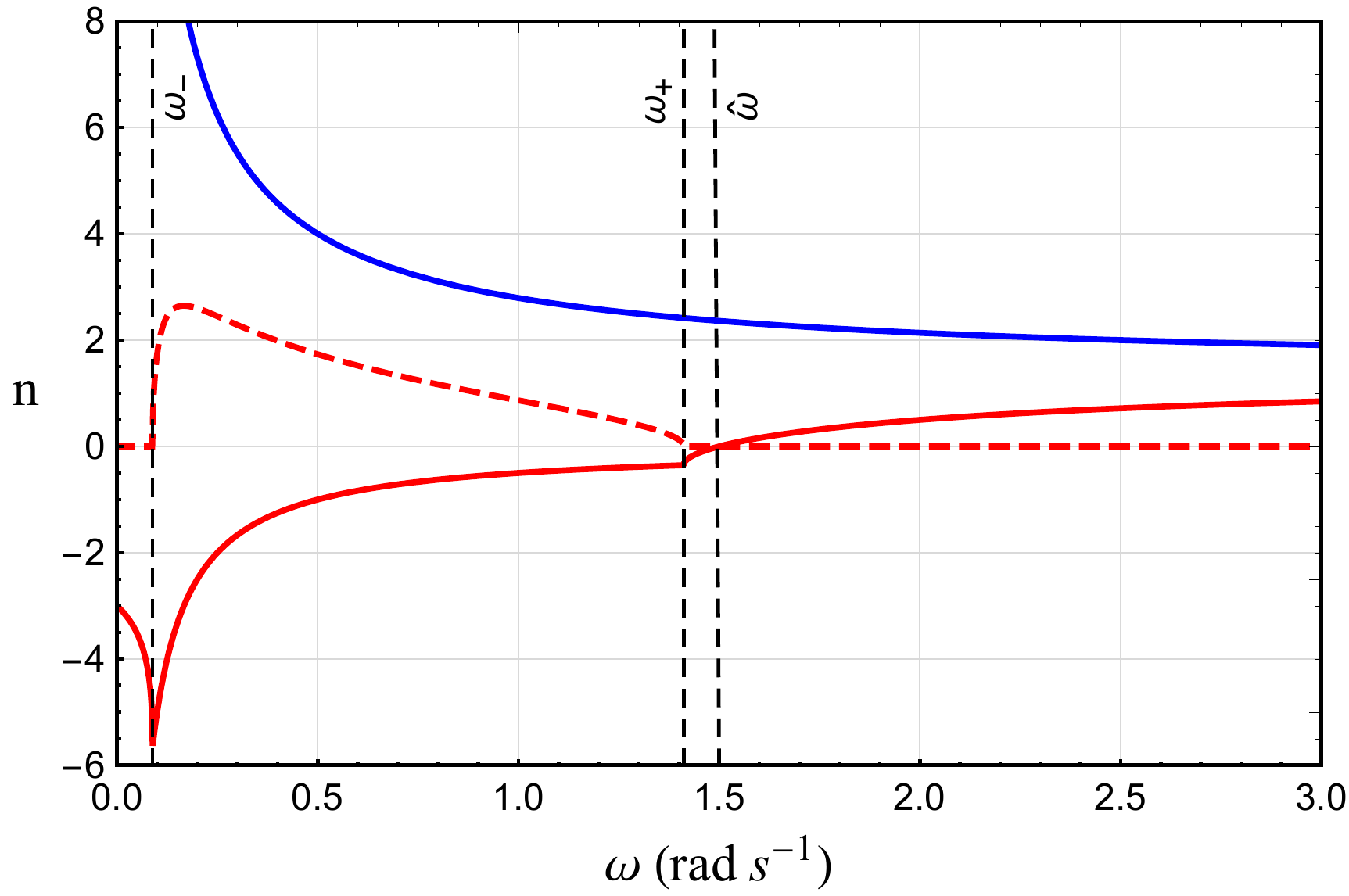}
	\caption{Refractive indices $n_{+}$ (red) and $\tilde{n}_{+}$ (blue). The solid (dashed) curves represent the real (imaginary) parts of the refractive indices. Here, we have used: $\mu=1$, $\epsilon=2$, $\Sigma_{H}=~3$~$\mathrm{s}^{-1}$, $\Sigma_{B}=-1$ $\mathrm{s}^{-1}$.}
	\label{plot-refraction-indices}
\end{figure}

\textbf{\textit{Dispersion relations}}. The chiral matter with CME and AHE can be effectively described by the Maxwell equations in continuous media, 
\begin{align}
\boldsymbol{\nabla} \cdot {\bf{D}} &= \rho, \label{maxwell-1} \\
\boldsymbol{\nabla} \times {\bf{H}} - \partial_{t} {\bf{D}} &= {\bf{J} }, \label{maxwell-2} \\
\boldsymbol{\nabla} \cdot {\bf{B}} &= 0, \label{maxwell-3} \\
\boldsymbol{\nabla} \times {\bf{E}} + \partial_{t} {\bf{B}} &= {\bf{0}}, \label{maxwell-4}
\end{align}
supplemented by the isotropic constitutive relations $\mathbf{D}=\epsilon\mathbf{E}$, $\mathbf{H}=\mu^{-1}\mathbf{B}$ (with $\mu$ and $\epsilon$ being constants) and the axionlike current density
\begin{align}
{\bf{J}} &= \sigma^{B} \cdot {\bf{B}} - \sigma^{H} \cdot ( \hat{{\bf{b}}} \times {\bf{E}} ), \label{general-magneticu-current-1}
\end{align}
with $\sigma^{B}$ and $\sigma^{H}$ being the magnetic and anomalous Hall conductivities. Implementing the plane wave \textit{ansatz}, one finds
\begin{align}
M_{ab} E^{b} &= 0, \label{Fresnel-1}
\end{align}
with
\begin{align}
M_{ab} &= n^{2} \delta_{ab} -n_{a}n_{b} - \omega^{2} \mu \bar{\epsilon_{ab}}, \label{Fresnel-2}
\end{align}
and  $k^{i} = \omega n^{i}$, 
\begin{align}
\bar{\epsilon}_{ab} &= \epsilon \delta_{ab} + \frac{i}{\omega} \varepsilon_{cjb} \left(\sigma^{B}_{ac}n_{j} - \sigma^{H}_{ac} \epsilon_{cjb} \hat{b}_{j} \right). \label{Fresnel-3}
\end{align}
Magnetic and Hall conductivity tensors, $\sigma^{B}$ and $\sigma^{H}$, are known to exhibit distinct parameterizations \cite{PedroPRD2020, Zelezny, Kurumaji}, which reflect different physical systems and symmetries. We here adopt the simple case of isotropic parametrizations, namely,
\begin{align}
\sigma^{B}_{ij} = \Sigma_{B} \delta_{ij}, \quad \sigma^{H}_{ij} = \Sigma_{H} \delta_{ij}, \label{super-reflectance-isotropic-case-1}
\end{align}
in such a way that the effective permittivity tensor reads
\begin{align}
\bar{\epsilon}_{ab} &= \epsilon\delta_{ab} - \frac{i}{\omega} \varepsilon_{abj} \left( \Sigma_{B} n_{j} - \Sigma_{H} \hat{b}_{j} \right), \label{super-reflectance-isotropic-case-2}
\end{align}
where $\Sigma_{B}$ and $\Sigma_{H}$ are constants. The non-trivial solutions of Eq.~(\ref{Fresnel-1}) requires $\mathrm{det}[M_{ij}]=0$, providing the following dispersion equation:
\begin{align}
0&=\left(n^{2} - \mu \epsilon \right)^{2} - n^{2} \frac{\mu^{2}}{\omega^{2}} \left( \Sigma_{B}^{2} - \frac{\Sigma_{H}^{2}}{\mu \tilde{\epsilon}} \sin^{2} \theta \right) + \nonumber \\
&\phantom{=} + \frac{\mu^{2}}{\omega^{2}}  \left(  2 n \Sigma_{B} \Sigma_{H} \cos\theta- \Sigma_{H}^{2} \right), \label{super-reflectance-isotropic-case-4}
\end{align}
with $\hat{\bf{n}} \cdot \hat{\bf{b}}= \cos \theta$, which yields the refractive indices resulting from the chiral effects already in consideration. Let us consider the scenario of propagation along $\hat{\bf{b}}$ direction, such that $\theta=0$. In this case, the refractive indices of the chiral matter are
\begin{align}
		n_{\pm} &= \frac{\mu \Sigma_{B}}{2\omega} \pm \sqrt{N_{-}(\omega) } ,\label{super-reflectance-isotropic-case-5} \\
		\tilde{n}_{\pm} &=- \frac{\mu \Sigma_{B}}{2\omega} \pm \sqrt{ N_{+}(\omega)} , \label{super-reflectance-isotropic-case-5-tilde-solution-1}
	\end{align}
where 
	\begin{equation}
		N_{\pm}(\omega) =\mu \epsilon + \left( \frac{\mu \Sigma_{B}}{2\omega} \right)^{2} \pm \frac{\mu \Sigma_{H}}{\omega}, \label{Npm}
	\end{equation}
	and we set $\Sigma_{H}>0$.
We are initially interested in the indices $n_{+}$ of \eqref{super-reflectance-isotropic-case-5} and $\tilde{n}_{+}$ of \eqref{super-reflectance-isotropic-case-5-tilde-solution-1}. The refractive indices $n_{-}$  and  $\tilde{n}_{-}$  will not be taken into account here, since they remain negative for all frequencies (for the conditions here adopted). While $\tilde{n}_{+}$ is always real and positive, the index $n_{+}$ presents distinct regimes depending on the frequency window. See Fig.~\ref{plot-refraction-indices}.   

To analyze the reflection properties, we will consider an interface between a simple dielectric characterized by constant electric permittivity $\epsilon_{1}$ and constant magnetic permeability $\mu_{1}$ (medium 1), defined in the region of $z<0$, and the chiral matter with refractive indices of \eqref{super-reflectance-isotropic-case-5} or \eqref{super-reflectance-isotropic-case-5-tilde-solution-1}, defined in the region $z>0$ (medium 2)\footnote{From this point onward, the substitutions $\epsilon \rightarrow \epsilon_{2}$ and $\mu \rightarrow \mu_{2}$ are applied to the previous expressions for $n_{+}$, $\tilde{n}_{-}$, $\hat{\omega}$, $\omega_{+}$, and $\omega_{-}$.
}. Considering usual boundary conditions at the interface and the normal pointing along $\hat{\bf{z}}$, one writes the reflection coefficients $R$ for $s$- and $p$-polarized incident wave, where $n_{2} =\{n_{+}, \tilde{n}_{+} \}$. For normal incidence, the reflection coefficient reads \cite{Jackson, Zangwill} 
\begin{align}
	R &= \left| \frac{ \mu_{2} \sqrt{\mu_{1} \epsilon_{1}} - \mu_{1} n_{2} }{  \mu_{2} \sqrt{\mu_{1} \epsilon_{1}} + \mu_{1}n_{2} } \right|^{2}.  \label{super-reflectance-isotropic-case-8}
\end{align}

\textbf{\textit{Modes associated with $n_{+}$}.} We initially examine situations of super-reflectance for the mode associated with the index $n_{+}$. Such an index has a rich behavior, being real or complex, positive or negative, depending on the frequency window. Considering $\Sigma_{B} <0$, one finds that $n_{+} $ becomes negative at $\omega <\hat{\omega}$, with
\begin{align}
	\hat{\omega}={ \Sigma_{H}}/{\epsilon}.\label{extra-3}
\end{align}
Additionally, $n_{+}$ becomes complex when $N_{-}(\omega)<0$, opening an absorption zone, $\omega_{-} < \omega < \omega_{+}$, with 
\begin{align}
	\omega_{\pm} &= \frac{\Sigma_{H}}{2\epsilon} \pm \frac{1}{2\epsilon} \sqrt{ \Sigma_{H}^{2} - \mu \epsilon \Sigma_{B}^{2} }, \label{super-reflectance-isotropic-case-21a}
\end{align}
where both roots are positive $(0< \omega_{-} < \omega_{+})$ and exist when
\begin{align}
	|\Sigma_{B}| \leq \frac{\Sigma_{H}}{ \sqrt{\mu \epsilon}} \, . \label{condition-absorption-band-1}
\end{align}
Furthermore, $n_{+}$ is real and negative for  $0 < \omega < \omega_{-}$ and $\omega_{+} < \omega < \hat{\omega}$, in accordance with the conditions  $N_{-} \geq 0$ and $\sqrt{ N_{-}(\omega)} < {\mu_{2} |\Sigma_{B}|}/{ 2\omega}$. See details in Fig.~\ref{plot-refraction-indices}.

\begin{figure}[h]
	\centering\includegraphics[scale=.6]{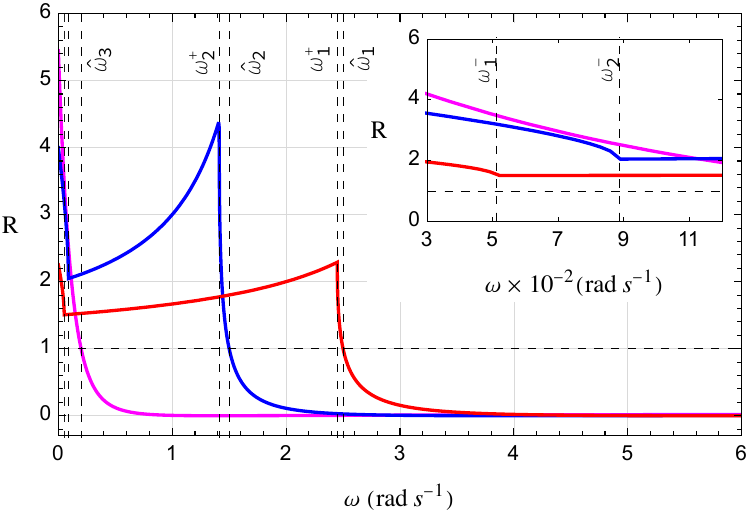}
	\caption{Reflectance at normal incidence of \eqref{super-reflectance-isotropic-case-8}. Here, we have used: $\mu_{1}=\mu_{2}=1$, $\epsilon_{1}=1$, $\epsilon_{2}=2$, and $\Sigma_{B}=-1$ for all curves. We have set $\Sigma_{H}= \{5, 3, 0.4\}$ $\mathrm{s}^{-1}$ for solid red, blue, and magenta lines, respectively. The dashed horizontal line indicates $R=1$. The dashed vertical lines, from left to right, are given by $\omega^{-}_{1}, \omega^{-}_{2}, \hat{\omega}_{3}, \omega^{+}_{2}, \hat{\omega}_{2}, \omega^{+}_{1}, \hat{\omega}_{1}$, where $\hat{\omega}_{i}$ and $\omega^{\pm}_{i}$ are given by \eqref{extra-3} and \eqref{super-reflectance-isotropic-case-21a}, respectively. The subscript $i$ indicates the examples plotted: $i=\{ 1, 2, 3\}$ for red, blue, and magenta curves. The inset plot highlights the behavior near the origin.}
	\label{plot-super-reflectance-examples}
\end{figure}

The behavior of $n_{+}$ yields reflectance greater than 1 in two frequency ranges, as examined below.
			\begin{itemize}
	\item Interval 1: where the refractive index $n_{+}$ is purely real and negative. As known, the index $n_{+}$ is real when $N_{-}(\omega)>0$, that is, $0<\omega<\omega_{-}$ and  $\omega> \omega_{+}$. In this case, one only has $R>1$ when $n_{+}$ is negative, implying 
		\begin{align}
			R &= \left| \frac{ \mu_{2} \sqrt{\mu_{1} \epsilon_{1}} + \mu_{1} \left|n_{+}\right|}{  \mu_{2} \sqrt{\mu_{1} \epsilon_{1}} - \mu_{1}\left|n_{+}\right| } \right|^{2}>1.  \label{super-reflectance-isotropic-case-8B}
		\end{align}	
Negative refraction ($n_{+} <0$) is obatined by writing
		\begin{align}
			n_{+} &= - \frac{\mu_{2} |\Sigma_{B}|}{2 \omega} +  \sqrt{ N_{-}(\omega) }, \label{super-reflectance-isotropic-case-9}
		\end{align}
for $\Sigma_{B}=-|\Sigma_{B}|$, and imposing the additional condition $\sqrt{ N_{-}(\omega)} < {\mu_{2} |\Sigma_{B}|}/{ (2\omega)}$, fulfilled for $\omega < \hat{\omega}$.
As $\hat{\omega}>\omega_{+}$, it turns out that the index $n_{+}$ is real and negative for
\begin{equation}
0 <\omega< {\omega}_{-}, \quad {\omega}_{+} <\omega< \hat{\omega},
\end{equation}
in which the associated wave at region 2 ($z>0$) propagates towards $(-\hat{\bf{z}})$ direction (negative refraction). As a result of this backward wave, the reflection coefficient becomes greater than unity, engendering the super-reflectance ($R>1$), as illustrated in Fig.~\ref{plot-super-reflectance-examples}.

\item Interval 2:  where the refractive index is complex, $n_{+}=n_{+}^{\prime}+in_{+}^{\prime\prime}$, with the real part negative, $n_{+}^{\prime}<0$, and imaginary piece positive,  providing 
\begin{align}
R = \left| \frac{ (\mu_{2} \sqrt{\mu_{1} \epsilon_{1}} + \mu_{1} \left|n_{+}^{\prime}\right|)^2+(n_{+}^{\prime\prime})^2}{ (\mu_{2} \sqrt{\mu_{1} \epsilon_{1}} - \mu_{1} \left|n_{+}^{\prime}\right|)^2+(n_{+}^{\prime\prime})^2 } \right|^{2}>1.  \label{super-reflectance-isotropic-case-8C}
\end{align}
The index $n_{+}$ becomes complex when $N_{-}(\omega)<0$, presenting non-null real (transmission) and imaginary (absorption) pieces. This happens for $\omega_{-} <\omega< \omega_{+}$, with $\omega_{\pm} $ given in \eqref{super-reflectance-isotropic-case-21a}. Differently from the previously examined situation, within this frequency window, the system exhibits reflectance that continuously increases with frequency and is greater than unity ($R>1$). Such a new anomalous behavior is depicted in Fig.~\ref{plot-super-reflectance-examples} and can be qualitatively explained as a consequence of $n_{+}=n_{+}^{\prime}+ia$, with negative real part, $n_{+}^{\prime}<0$. As the refractive index $n_{+}$ becomes complex, the wavevector reads
\begin{equation}
k_{+} = \mu_{2}\Sigma_{B} / 2 + i a,
\end{equation}
with $a=\sqrt{ -N_{-}(\omega)}$ being a positive real number. For $\Sigma_{B} >0$, the usual plane wave solution ${\bf{E}} = {\bf{E}}_{0} e^{i [ k_{+} (\hat{\bf{z}} \cdot {\bf{r}}) - \omega t ] } $  yields an exponentially decreasing amplitude wave in medium 2 in its propagation direction ($+\hat{z}$), 
\begin{align}
	e^{i [ k_{+} (\hat{\bf{z}} \cdot {\bf{r}}) - \omega t ]}= e^{-a (\hat{\bf{z}} \cdot {\bf{r}})} \, e^{i \left(\frac{\mu}{2} |\Sigma_{B}| (\hat{\bf{z}} \cdot {\bf{r}}) - \omega t\right)}. \label{super-reflectance-isotropic-case-18}
\end{align}
However,  for $\Sigma_{B} = - | \Sigma_{B}|$, the plane wave solution in medium 2 becomes
\begin{equation}
	{\bf{E}} = {\bf{E}}_{0} \, e^{a [\,(-\hat{\bf{z}}) \cdot {\bf{r}}]} \, e^{i \left\{ \frac{\mu}{2} |\Sigma_{B}| \, \left[ (-\hat{\bf{z}}) \cdot {\bf{r}}\right] - \omega t \right\}   } , \label{super-reflectance-isotropic-case-20}
\end{equation}
engendering a wave with exponential growth towards the ``negative" propagation direction $(-\hat{\bf{z}})$, which contributes magnifying the reflection coefficient $R$ (back to medium 1).

\end{itemize}

Note that for $\omega > \hat{\omega}$, the index $n_{+}$ becomes real and positive even for $\Sigma_{B}<0$, see \eqref{super-reflectance-isotropic-case-5}, impling in $R<1$ (usual reflectance), as expected. The aspects of anomalous reflectance associated with the index $n_{+}$, for $\Sigma_{B} < 0$ and $\Sigma_{H} > 0 $, are illustrated in Fig.~\ref{plot-super-reflectance-examples} for three sets of parameters values, where one observes three disctint windows for super reflectance: $0 <\omega< {\omega}_{-}$,  $\omega_{-} <\omega< \omega_{+}$, ${\omega}_{+}<\omega<\hat{\omega}$. 

The scenario depicted in Fig. \ref{plot-super-reflectance-examples} takes place when the two conductivities exhibit opposite signs, $\Sigma_B<0$ and  $\Sigma_H>0$.

\begin{figure}[H]
	\centering\includegraphics[scale=.6]{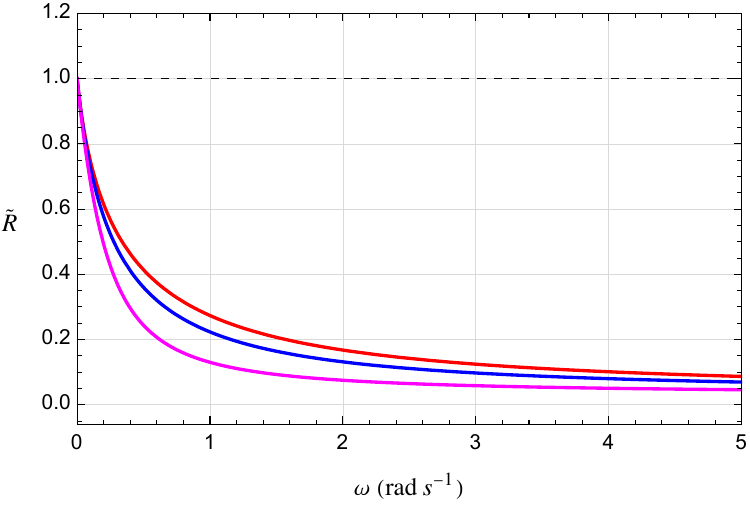}\caption{Reflectance at normal incidence of \eqref{tilde-refractive-index-3}. Here, we have used: $\mu_{1}=\mu_{2}=1$, $\epsilon_{1}=1$, $\epsilon_{2}=2$, and $\Sigma_{B}=-1$ for all curves. We have set $\Sigma_{H}= \{5, 3, 0.4\}$ $\mathrm{s}^{-1}$ for solid red, blue, and magenta lines, respectively. The dashed horizontal line indicates $R=1$.}
	\label{plot-super-reflectance-n-tilde-examples}
\end{figure}

\textbf{\textit{{Modes associated with $\tilde{n}_{+}$}}.} From \eqref{super-reflectance-isotropic-case-5-tilde-solution-1}, we observe that $\tilde{n}_{+}$ is real and positive for all frequencies, regardless of the sign of $\Sigma_{B}$. In this case, there is no negative refraction. Thus, no anomalous reflectance is expected. Indeed, at medium 2, one will only observe a usual electromagnetic wave propagating at the $(+ \hat{\bf{z}})$ direction without attenuation. The reflectance for the propagating mode related to $\tilde{n}_{+}$ is given by
\begin{align}
\tilde{R_+} &= \left| \frac{ \mu_{2} \sqrt{\mu_{1} \epsilon_{1}} - \mu_{1} \tilde{n}_{+} }{  \mu_{2} \sqrt{\mu_{1} \epsilon_{1}} + \mu_{1}\tilde{n}_{+} } \right|^{2},  \label{tilde-refractive-index-3}
\end{align}
being always smaller than 1, as depicted in Fig.~\ref{plot-super-reflectance-n-tilde-examples}.  This happens because the negative refraction ($\tilde{n}_{+} < 0$) does not occur for $\omega>0$ (with $\Sigma_{B} <0$ or $\Sigma_{B}>0$).

\textbf{\textit{Very low-frequency regime.}} Since the magnetic conductivity and Hall term introduce a $\omega^{-1}$ dependence in the refractive indices, it is natural to investigate the low-frequency regime, where these contributions can be significant. Thus, in the limit of very-low frequencies, the refractive index $n_{+}$ (or $k_{+}=\omega n_{+}$), given in \eqref{super-reflectance-isotropic-case-5} behaves as
\begin{align}
n_{+} \simeq \frac{\mu_{2} |\Sigma_{B}|}{2\omega} \left[1+\mathrm{sgn}(\Sigma_{B}) \right] - \frac{ \Sigma_{H}   }{|\Sigma_{B}|} . \label{small-frequency-n-plus-3}
\end{align}
The reflection coefficient in the small frequency regime is written now as
\begin{align}
R= \left| \frac{  \mu_{2} \sqrt{\mu_{1} \epsilon_{1}} - \frac{ \mu_{1}\mu_{2}}{2\omega} |\Sigma_{B}| \, [1 + \mathrm{sgn}(\Sigma_{B})  ] + \mu_{1} \frac{ (\Sigma_{H} )}{ |\Sigma_{B}| } } {\mu_{2} \sqrt{\mu_{1}\epsilon_{1}} + \frac{ \mu_{1}\mu_{2}}{2\omega}  |\Sigma_{B}|  \,  [1+ \mathrm{sgn}(\Sigma_{B}) ] - \mu_{1} \frac{ (\Sigma_{H} )}{ |\Sigma_{B}|} } \right|^{2} , \label{small-frequency-R-1} \\
\end{align} 
For $\Sigma_{B}<0$, the latter expression provides
\begin{align}
\left. \frac{}{} R\right|_{\Sigma_{B}<0} &=  \frac{ \left( \mu_{2} \sqrt{\mu_{1}\epsilon_{1}} |\Sigma_{B}| + \mu_{1} \Sigma_{H} \right)^{2} } { \left( \mu_{2} \sqrt{\mu_{1}\epsilon_{1}} |\Sigma_{B}| - \mu_{1} \Sigma_{H} \right)^{2} }  , \label{small-frequency-R-3}
\end{align}
which yields $R>1$. This result is frequency independent and corresponds to the zero-frequency limit of the examples of Fig.~\ref{plot-super-reflectance-examples}. It also recovers the outcomes of Ref.~\cite{Nishida2} (with $\sigma=0$). For $\Sigma_{B} >0$, equation (\ref{small-frequency-R-1}) yields 
\begin{align}
  \left. \frac{}{} R\right|_{\Sigma_{B}>0} &=  \frac{ \left\{ -2 \mu_{1} \mu_{2} \Sigma_{B}^{2} + 2 \omega \left[ \mu_{2} \sqrt{\mu_{1} \epsilon_{1}} \Sigma_{B} + \mu_{1} \Sigma_{H} \right] \, \right\}^{2} } {  \left\{ 2 \mu_{1} \mu_{2} \Sigma_{B}^{2} + 2\omega \left[ \mu_{2} \sqrt{\mu_{1} \epsilon_{1}} \Sigma_{B} - 2\mu_{1} \Sigma_{H} \right] \, \right\}^{2}  }, \label{small-frequency-R-4}
  \end{align}
which, in the limit of zero frequency, provides $\lim_{\omega \rightarrow 0} \left. \frac{}{} R\right|_{\Sigma_{B}>0} =1$, being the usual result obtained in the absence of negative refraction.

A similar low-frequency analysis for the index $\tilde{n}_{+}$ shows that the reflection coefficient is always less than unity,
\begin{align}
\lim_{\omega \rightarrow 0} \left. \tilde{R} \right|_{\Sigma_{B}>0} &=  \frac{ \left( \mu_{1} \Sigma_{H} - \mu_{2} \sqrt{\mu_{1} \epsilon_{1} } \Sigma_{B} \right)^{2} } { \left( \mu_{1} \Sigma_{H} + \mu_{2} \sqrt{\mu_{1} \epsilon_{1}} \Sigma_{B} \right)^{2} } ,  \label{small-frequency-R-tilde-3}
\end{align}
or equal to 1, 
\begin{align}
\lim_{\omega \rightarrow 0} \left. \tilde{R} \right|_{\Sigma_{B} < 0} &= 1,  \label{small-frequency-R-tilde-4}
\end{align}
case that corresponds to the zero-frequency observed in Fig.~\ref{plot-super-reflectance-n-tilde-examples}.

\textbf{\textit{Complex Kerr rotation.}} In this section, we discuss another optical route to determine the relative sign between the magnetic and Hall conductivities, using reflection signatures of the axion dielectric with the AHE term. To to this, one analyzes the Kerr rotation angle and Kerr ellipticity angle, obtained by considering incident light from medium 1 (simple matter with $\epsilon_{1}$ and $\mu_{1}$) and medium 2 (axion dielectric with $\mu_{2}$, $\epsilon_{2}$, $\Sigma_{B}$ and $\Sigma_{H}$).

Considering the refractive indices, $n_{+}$ and $\tilde{n}_{+}$, which are associated with LCP and RCP waves, respectively, the corresponding reflection coefficients are given by
\begin{subequations}
\begin{align}
	r &= \frac{ \mu_{2} \sqrt{\mu_{1} \epsilon_{1}} - \mu_{1} n_{+} }{  \mu_{2} \sqrt{\mu_{1} \epsilon_{1}} + \mu_{1}n_{+} } , \label{anomalous-kerr-1} \\
	\tilde{r} &= \frac{ \mu_{2} \sqrt{\mu_{1} \epsilon_{1}} - \mu_{1} \tilde{n}_{+} }{  \mu_{2} \sqrt{\mu_{1} \epsilon_{1}} + \mu_{1}\tilde{n}_{+} } . \label{anomalous-kerr-2}
\end{align}
\label{kerr}
\end{subequations}

The complex Kerr rotation is characterized by the Kerr rotation angle, $\theta_{K}$, and the ellipticity angle, $\eta_{K}$, given by
\begin{align}
\tan(2\theta_{K}) &= - \frac{ 2 \,  \mathrm{Im} \,( \Delta) }{1 - \left| \Delta \right|^{2}} , \label{anomalous-kerr-3} \\
\sin( 2\eta_{K})  &= \frac{ 2 \, \mathrm{Re} \, ( \Delta) }{1 + \left| \Delta \right|^{2} } , \label{anomalous-kerr-4}
\end{align}
with
\begin{align}
\Delta &= \frac{r - \tilde{r}}{r+ \tilde{r}} . \label{anomalous-kerr-5}
\end{align}
As known, the sign of $\eta_{K}$ indicates the handedness of the reflected wave, that is left- (right-) handed elliptically polarized for so that $\eta_{K} >0$ ($\eta_{K}<0$), respectively \cite{Zangwill}.

Using now \eqref{anomalous-kerr-1} and \eqref{anomalous-kerr-2}, one finds
\begin{align}
\Delta &= - \frac{ \mu_{2} \sqrt{\mu_{1} \epsilon_{1}} \left( \frac{\mu_{2} \Sigma_{B}}{\omega} + \sqrt{N_{-}} - \sqrt{N_{+}} \right) } { \mu_{2}^{2} \epsilon_{1} + \mu_{1} \left( \frac{\mu_{2} \Sigma_{B}}{2\omega} + \sqrt{N_{-}} \right) \, \left( \frac{\mu_{2} \Sigma_{B}}{2\omega} - \sqrt{N_{+}} \right) } , \label{anomalous-kerr-6-extra-1}
\end{align}
where
\begin{align}
N_{\pm} &= \mu_{2}\epsilon_{2} +\left( \frac{\mu_{2}\Sigma_{B}}{2\omega}\right)^{2} \pm \frac{\mu_{2}\Sigma_{H}}{\omega} . \label{anomalous-kerr-6-extra-2} 
\end{align}

Notice that \eqref{anomalous-kerr-5} becomes complex when $N_{-}(\omega) <0$, 
which happens within the frequency band $\omega_{-} <\omega< \omega_{+}$, with $\omega_{\pm}$ given in \eqref{super-reflectance-isotropic-case-21a}. Considering this scenario, one can rewrite \eqref{anomalous-kerr-6-extra-1} as
\begin{align}
\left. \Delta \right|_{\Delta\omega \neq 0} &= \Delta' + i \Delta'', \label{anomalous-kerr-7}
\end{align}
where
\begin{align}
\Delta' &= f \left[ \left( \frac{\mu_{2}}{\omega} \Sigma_{B} - \sqrt{N_{+}} \right) A + \mu_{1} P^{3/2} \left( \frac{\mu_{2} }{2\omega}\Sigma - \sqrt{N_{+}} \right) \right],  \label{anomalous-kerr-8} \\
\Delta^{\prime \prime} &= f \left[ A P - \mu_{1} \sqrt{P} \left(\frac{\mu_{2} \Sigma_{B}}{2\omega} - N_{+} \right) \right], \label{anomalous-kerr-9}
\end{align}
with
\begin{align}
A &= \mu_{2}^{2} \epsilon_{1} + \mu_{1} \left[  \left(\frac{\mu_{2} \Sigma_{B}}{2\omega} \right)^{2} - \frac{\mu_{2} \Sigma_{B} }{2\omega} \sqrt{N_{+}} \right] , \label{anomalous-kerr-10} \\
P&= \frac{\mu_{2} \Sigma_{H} }{\omega} - \left( \frac{\mu_{2} \Sigma_{B}}{2\omega} \right)^{2} - \mu_{2} \epsilon_{2} , \label{anomalous-kerr-11} \\
f &= - \frac{\mu_{2} \sqrt{\mu_{1} \epsilon_{1} }} { A^{2} + \mu_{1}^{2} P \left( \frac{ \mu_{2}\Sigma_{B}}{2\omega} - \sqrt{N_{+}} \right)^{2}} .  \label{anomalous-kerr-12} 
\end{align}

\begin{figure}[h]
	\centering\includegraphics[scale=.6]{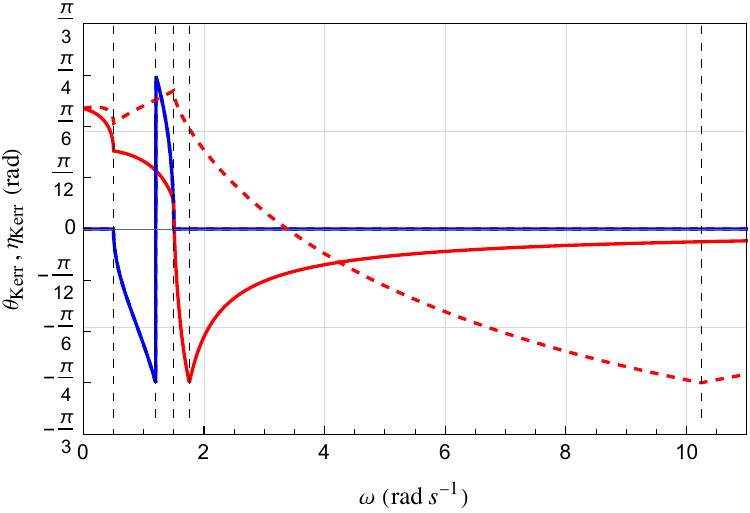}
	\caption{Kerr rotation, $\theta_{K}$, and ellipticity, $\eta_{K}$, angles, given by of Eqs.~(\ref{anomalous-kerr-3}) and (\ref{anomalous-kerr-4}), respectively, for $|\Sigma_{B}| < \Sigma_{H} / \sqrt{\mu_{2}\epsilon_{2}}$. The blue line represents $\theta_{K}$, endowed with giant Kerr rotation. The red curves illustrate $\eta_{K}$ for $\Sigma_{B}>0$ (solid red) and $\Sigma_{B} <0$ (dashed red). Here, we have used: $\mu_{1}=\mu_{2}=1$, $\epsilon_{1}=2$, $\epsilon_{2}=3$, $\Sigma_{B}=-3$ (dashed red line), $\Sigma_{B}=3$ (solid red line), and $\Sigma_{H}=6$ $s^{-1}$. The dashed vertical lines (from left to right) are given by, respectively: $\omega_{-}$, $\omega'$, $\omega_{+}$, $\omega^{+}_{-\pi/4}$, and $\omega^{-}_{-\pi/4}$, with $\omega_{\pm}$ of \eqref{super-reflectance-isotropic-case-21a}, $\omega'$ is obtained by \eqref{omegalinha}, and $\omega^{\pm}_{-\pi/4}$ are given by \eqref{frequencies-for-Kerr-ellipticity-equals-minus-pi-over-4-equation-1}. Note that the two frequencies in (\ref{frequencies-for-Kerr-ellipticity-equals-minus-pi-over-4-equation-1}) do not appear in the same curve.}
	\label{plot-Kerr-rotation-and-ellipticity}
\end{figure}

\begin{figure}[H]
	\centering\includegraphics[scale=.6]{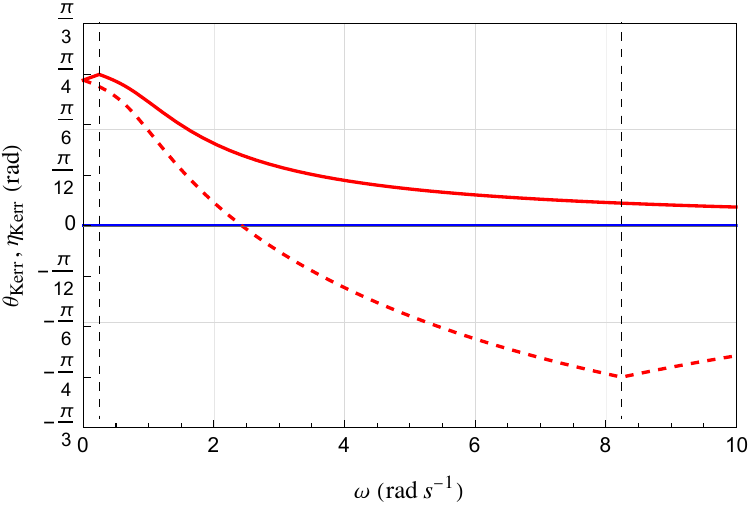}
	\caption{Kerr rotation, $\theta_{K}$, and ellipticity, $\eta_{K}$, angles, given by of Eqs.~(\ref{anomalous-kerr-3}) and (\ref{anomalous-kerr-4}), respectively, for $|\Sigma_{B}| > \Sigma_{H} / \sqrt{\mu_{2}\epsilon_{2}}$. The blue (red) curves illustrates $\theta_{K}$ ($\eta_{K}$) for $\Sigma_{B}>0$ (solid red) and $\Sigma_{B} <0$ (dashed red). Here, we have used: $\mu_{1}=\mu_{2}=1$, $\epsilon_{1}=2$, $\epsilon_{2}=3$, $\Sigma_{B}=-3$ (dashed red line), $\Sigma_{B}=3$ (solid red line), and $\Sigma_{H}=4$ $s^{-1}$. The dashed vertical lines (from left to right) are given by, respectively: $\omega^{+}_{+\pi/4}$ and $\omega^{-}_{-\pi/4}$. }
	\label{plot-Kerr-rotation-and-ellipticity-without-discontinuity}
\end{figure}

To illustrate the general behavior of Kerr rotation and Kerr ellipticity, we depict $\theta_{K}$ and $\eta_{K}$ in Fig.~\ref{plot-Kerr-rotation-and-ellipticity} and Fig.~\ref{plot-Kerr-rotation-and-ellipticity-without-discontinuity}, highlighting that the condition (\ref{condition-absorption-band-1}) does not hold in Fig.~\ref{plot-Kerr-rotation-and-ellipticity-without-discontinuity}.

Regarding Fig.~\ref{plot-Kerr-rotation-and-ellipticity}, one can determine the frequency window, $\Delta \omega = \omega_{+} - \omega_{-}$, where the Kerr rotation angle $\theta_{K}$ is non null, with $\omega_{\pm}$ given by \eqref{super-reflectance-isotropic-case-21a}. This is the range where $\Delta$ becomes complex.  Additionally, such rotation angle can exist provided that the relation $ |\Sigma_{B}| < \Sigma_{H}/\sqrt{\mu_{2}\epsilon_{2}}$ holds. Giant Kerr rotation ($\theta_{K}=\pm \pi/4$) can occur at frequency $\omega^{\prime}$ where $\theta_{K}$ abruptly changes sign\footnote{Giant Kerr was also reported in the context of Weyl semimetals \cite{Sonowal}.}. Such a frequency  can be found by setting the denominator of \eqref{anomalous-kerr-3} equal to zero, that is,
\begin{equation}
\left[1-\Delta^{\prime 2} - \Delta^{\prime \prime 2}\right]_{\omega=\omega^{\prime}}=0.
\label{omegalinha}
\end{equation} 
Furthermore, the Kerr angle signals behave with the frequency in a highly nontrivial way, especially the Kerr ellipticity, which may have (or not) maximum values of $\pm \pi/4$. This happens due to the large number of electromagnetic parameters in the system and also to the structure of the $N$ function, which is quadratic in the frequency. Consequently, the profiles of Kerr ellipticity can vary significantly from one particular case to another. In general, the values of $\eta_{K}=\pm \pi/4$ can occur whenever the expression (\ref{anomalous-kerr-4}) can be equal to $\pm 1$. Considering the particular scenario $\epsilon_{2} >\epsilon_{1}$ and $\mu_{1}=\mu_{2}$, one finds that $\eta_{K}=-\pi/4$ rad happens at frequencies $\omega^{\pm}_{-\pi/4}$, given by
\begin{align}
\omega^{\pm}_{-\pi/4} &= -\mathrm{sign}[\Sigma_{B}]  \frac{\mu_{2} |\Sigma_{B}| \sqrt{\mu_{1} \epsilon_{1}}}{|\mu_{1}\epsilon_{2} - \mu_{2} \epsilon_{1} |} + \frac{\mu_{1} \Sigma_{H}}{\mu_{1} \epsilon_{2} - \mu_{2}\epsilon_{1}},  \label{frequencies-for-Kerr-ellipticity-equals-minus-pi-over-4-equation-1}
\end{align}
where the superscript $\pm$ is connected to the sign of $\Sigma_{B}$ (it holds for $\Sigma_{B}$ positive and negative). On the other hand, $\eta_{K} = + \pi/4$ arises at frequency,
\begin{align}
\omega^{+}_{+\pi/4} &= \frac{\mu_{2} |\Sigma_{B}| \sqrt{\mu_{1}\epsilon_{1}}}{| \mu_{1} \epsilon_{2}-\mu_{2} \epsilon_{1}|} - \frac{\mu_{1} \Sigma_{H}}{\mu_{1}\epsilon_{2} - \mu_{2} \epsilon_{1}} ,\label{frequency-for-Kerr-ellipticity-equal-pi-over-4-1}
\end{align}
which holds when $|\Sigma_{B}| > \Sigma_{H} / \sqrt{\mu_{2}\epsilon_{2}}$ and  $\Sigma_{B}>0$.

In Figs.~\ref{plot-Kerr-rotation-and-ellipticity} and \ref{plot-Kerr-rotation-and-ellipticity-without-discontinuity}, the Kerr rotation angle ($\theta_{K}$) is described by solid (dashed) blue lines for $\Sigma_{B} >0$ ($\Sigma_{B} <0$). Although $\theta_{K}$ for both cases lie on top of each other in the examples illustrated, they are actually distinct by a tiny quantity, so that such a difference will not be here taken into account\footnote{They differ as 
		$\Delta|_{\Sigma_{B}>0}  - \Delta|_{\Sigma_{B}<0} = - \frac{ \mu_{2} |\Sigma_{B} | }{\omega} \frac{ \sqrt{\mu_{1} \epsilon_{1} } (\mu_{2} \epsilon_{1} - \mu_{1} \epsilon_{2} ) } { M  } $, 
		with
		$M= (1/2) (\mu_{2}^{2} \epsilon_{1}^{2} + \mu_{1}^{2} \epsilon_{2}^{2} ) - \mu_{1}^{2} \Sigma_{H}^{2} / (2 \omega^{2} ) + \mu_{1} \epsilon_{1} N$, and $N= \left( \frac{\mu_{2} \Sigma_{B} }{2\omega} \right)^{2} - \sqrt{N_{-}} \sqrt{N_{+}}$. However, in the example illustrated in Figs.~\ref{plot-Kerr-rotation-and-ellipticity} and \ref{plot-Kerr-rotation-and-ellipticity-without-discontinuity} the difference in $\theta_{K}$ for $\Sigma_{B} > 0$ and $\Sigma_{B}<0$ is too small ($ \sim 10^{-15}$ rad) to be considered.}. It means that the sign of $\Sigma_B$ can not be read by the  angle $\theta_{K}$, but can be determined by the behavior of $\eta_{K}$, as shown in  Figs.~\ref{plot-Kerr-rotation-and-ellipticity} and \ref{plot-Kerr-rotation-and-ellipticity-without-discontinuity}.

\textbf{\textit{Final Remarks}.} In this work, we have investigated optical effects in systems endowed with magnetic conductivity and anomalous Hall term, as Weyl semimetals. Particularly, we have addressed reflection properties considering an interface between a simple dielectric and a material with both magnetic and Hall conductivities, addressing the anomalous reflectance. The reflectance greater than unity occurs when $\Sigma_{B}<0$ and $\Sigma_{H}>0$. In this scenario, the emergence of negative refraction in the system occurs for $0 <\omega< {\omega}_{-}$ and  ${\omega}_{+} <\omega< \hat{\omega}$, with $\hat{\omega}$ of \eqref{extra-3}, and makes the electromagnetic wave in the medium 2 (characterized by $\Sigma_{B}$ and $\Sigma_{H}$) to propagate back to medium 1. Such unusual behavior enhances the amplitude of the reflected wave \cite{Nishida2, Alex2024}, yielding $R>1$. Additionally, for $|\Sigma_{B}|<\Sigma_{H} / \sqrt{\mu_{2}\epsilon_{2}}$, one finds a frequency window $\Delta \omega = \omega_{+} - \omega_{-}$, with $\omega_{\pm}$ of \eqref{super-reflectance-isotropic-case-21a},  where the reflectance also exhibits a frequency-dependent increasing behavior, fed by the exponential amplication exhibited in Eq.~(\ref{super-reflectance-isotropic-case-20}), which highlights the signature of this anomalous optical phenomenum. Effectively, such an enhancement happens when the real and imaginary pieces of the refractive index exhibit opposite signs.

Another optical signature of the system ruled by the axion electrodynamics stems from the complex Kerr rotation at normal incidence. For $|\Sigma_{B}| < \Sigma_{H} / \sqrt{\mu_{2}\epsilon_{2}}$, one has $\theta_{K}\neq 0$ in the absorption window,  $\omega_{-} <\omega < \omega_{+}$, range in which giant Kerr rotation takes place, see Fig.~\ref{plot-Kerr-rotation-and-ellipticity}. As for the ellipticity, it assumes the minimum value $-\pi/4$ for $\Sigma_{B} < 0$ and $\Sigma_{B} > 0$. The maximum value $+\pi/4$ is not achieved in this case. 

For $|\Sigma_{B}| > \Sigma_{H} / \sqrt{\mu_{2}\epsilon_{2}}$, the absorption frequency window desappears, $\Delta \omega = 0$, leading to a null Kerr rotation ($\theta_{K}=0$), while the Kerr ellipticity, $\eta_{K}$, varies smoothly with the frequency (see Fig.	\ref{plot-Kerr-rotation-and-ellipticity-without-discontinuity}) and plays a key role for determing the sign of the magnetic conductictivity: for $\Sigma_{B} >0$ one has the curve with maximum, $\eta_{K} = + \pi/4$ rad, while for $\Sigma_{B} <0$ it corresponds the dashed curve with minimum, $\eta_{K} = -\pi/4$ rad.

Although reflectance greater than unity seems to violate energy conservation, it is important to mention that in Weyl semimetals with both chiral magnetic conductivity and AHE, the chiral imbalance drives the system out of equilibrium. Relaxation towards the equilibrium can supply the additional energy involved in the anomalous reflectance \cite{Nishida2}. This unstable behavior stems from both CME and the AHE conductivities, whose joint effect drives a nonequilibrium configuration. Thus, the additional energy for the anomalous reflected wave ($R >1$)  can be interpreted as a consequence of the unstable electromagnetic waves in a nonequilibrium state. Such effects were also addressed in Ref. \cite{Alex2024}.

\textit{\textbf{Acknowledgments}}.  M.M.F. is supported by FAPEMA APP-12151/22, CNPq/Produtividade 311220/2019-3, and CNPq/Universal/422527/2021-1. P.D.S.S. thanks FAPEMA APP-12151/22. We are indebted to CAPES/Finance Code 001 and FAPEMA/POS-GRAD-04755/24.

\end{document}